\documentclass[10pt,showpacs,twocolumn]{revtex4}
\usepackage{graphicx}
\usepackage{amsmath}
\usepackage{amssymb}
\usepackage{multirow}
\usepackage{}

\begin{document}
\title{Quenching effect in below-threshold high harmonic generation}
\author{Xiaosong Zhu,$^{1}$ Xi Liu,$^{1}$ Pengfei Lan,$^{1}$ \footnote{Corresponding author: pengfeilan@hust.edu.cn} Qingbin Zhang,$^{1}$ Dian Wang,$^{1}$ Yueming Zhou,$^{1}$ Min Li,$^{1}$ and Peixiang Lu$^{1,2}$ \footnote{Corresponding author: lupeixiang@hust.edu.cn}}

\affiliation{$^1$ School of Physics and Wuhan National Laboratory for Optoelectronics, Huazhong University of
Science and Technology, Wuhan 430074, China\\
$^2$ Laboratory of Optical Information Technology, Wuhan Institute of Technology, Wuhan 430205, China}
\date{\today}

\begin{abstract}
We theoretically demonstrate the quenching effect in below-threshold high harmonic generation (HHG) by using the time-dependent density-functional theory (TDDFT) and solving the time-dependent Schr\"{o}dinger equation (TDSE). It is shown that the HHG is substantially suppressed in particular harmonic orders in the below-threshold region when multi-electron interaction comes into play. The position of the suppression is determined by the energy gap between the highest occupied orbital and the higher-lying orbital of the target. We show that the quenching effect is due to a new class of multi-electron dynamics involving electron-electron energy transfer, which is analog to the fluorescence quenching owing to the energy transfer between molecules in fluorescent material. This work reveals the important role of the multi-electron interaction on HHG especially in the below-threshold region.
\end{abstract}  \maketitle

High harmonic generation (HHG) is a highly nonlinear process arising when atoms or molecules are exposed to intense femtosecond laser pulses. Investigation on HHG has received a lot of concerns because of two fascinating applications: the generation of attosecond pulses in the extreme ultraviolet and soft X-ray region \cite{Mairesse,Goulielmakis} and the self-probing of atoms and molecules from the high harmonic spectrum  \cite{Lein,Haessler}. The high harmonic spectroscopy allows an unprecedented combination of attosecond and {\AA}ngstr\"{o}m resolutions, providing a powerful tool to gain insight into the microscopic world \cite{Itatani,Baker,Liwen,Smirnova,Bertrand,Wong,Cireasa}.

Most previous works focus on the high harmonics above the ionization threshold, i.e. the harmonics with photon energy higher than the ionization potential of the target. Recently, increasing attentions are paid to the below-threshold HHG, due to the potential application of generating extreme ultraviolet frequency combs for the precision spectroscopy and metrology \cite{Cingoz,Chini}. However, compared with the HHG above threshold, the process of below-threshold HHG is much more complex. The mechanism of below-threshold HHG has been much less explored and many questions still remain debated \cite{Xiong,Lipengcheng}.

Moreover, in most studies on HHG, single active electron (SAE) approximation is routinely adopted. Within this approximation, the HHG process is understood by the behavior of the single active electron and the roles of other electrons are ignored. It is generally considered that the theories relying on SAE approximation can well explain most of the characteristics of HHG, especially for atomic targets. For molecular systems, since the energies of different orbitals are much closer, it was shown that the contribution from lower-lying orbitals to HHG plays a nonnegligible role \cite{Smirnova,Worner,Haessler2}. However, the effect of dynamical electron-electron ($e$-$e$) interactions on HHG has seldom been investigated, except for only a few works \cite{Gordon,Shiner,Pabst}. It was shown that the collective multi-electron dynamics lead to a characteristic giant resonant enhancement of HHG near the cutoff region in the high harmonic spectrum of Xe \cite{Shiner,Pabst}.

In this work, we demonstrate a quenching effect in below-threshold HHG using the time-dependent density-functional theory (TDDFT) and solving the time-dependent Schr\"{o}dinger equation (TDSE). By comparing the harmonic spectra from different systems with one or multiple active electrons, we show that this quenching effect is a result of the multi-electron interaction. It is also found that the position where the quenching occurs is determined by the energy gap between the highest occupied orbital and the higher-lying orbital of the target. The quenching effect is attributed to a new class of multi-electron dynamics involving $e$-$e$ energy transfer. This dynamics can be verified from the anomalous ellipticity dependence of HHG at the quenched harmonics.

The response of multi-electron systems in strong laser fields are simulated by using the three-dimensional (3D) TDDFT \cite{Runge}. The evolutions of the systems follow the time-dependent Kohn-Sham (TDKS) equations (atomic units are used unless otherwise stated)
\begin{equation}
i\frac{\partial}{\partial t}\Psi_i(\mathbf{r}, t)=[-\frac{\nabla^2}{2}+v_{s}(\mathbf{r},t)]\Psi_i(\mathbf{r}, t),  \quad  i=1,2,\cdots,N.
\end{equation}
$N$ is the number of Kohn-Sham (KS) orbitals $\Psi_i(\mathbf{r}, t)$, and we do not specify spin-orbitals. In addition, we freeze the electrons in the inner shells and only the electrons in the valence shell are active. $v_s(\mathbf{r}, t)$ is the KS potential, which is defined as
\begin{equation}
v_{s}(\mathbf{r},t)=\int{\frac{\rho({\mathbf{r'},t})}{|\mathbf{r}-\mathbf{r'}|}}d\mathbf{r'}+v_{xc}({\mathbf{r},t})+v_{ne}+\mathbf{E}(t)\cdot\mathbf{r}.
\end{equation}
$\rho({\mathbf{r},t})$ is the time-dependent density of the multi-electron interacting system given by $\rho({\mathbf{r},t})=\sum_{i=1}^N{|\Psi_i(\mathbf{r},t)|^2}$. $v_{xc}$ is the exchange-correlation potential including all non-trivial many body effects. Here, we apply the well-known optimized effective potential (OEP) in Krieger-Li-Iafrate (KLI) approximation \cite{OEP-KLI}. $v_{ne}$ is the ionic potential modeled by the norm-conserving nonlocal Troullier-Martines pseudopotential \cite{psf}. $\mathbf{E}(t)$ is the electric field of the laser. Before propagation in time, the initial states are obtained by solving the KS equations self-consistently at the density functional theory (DFT) level. All the simulations in this work are performed with \cite{Octopus}. The harmonic spectrum is obtained as $S(\omega)=|\int{\ddot{\mathbf{d}}(t)\exp(-i\omega t)}dt|^2$, where the time-dependent dipole is calculated as $\mathbf{d}(t)=\int{\mathbf{r}\rho(\mathbf{r},t)}d\mathbf{r}$.

We take the commonly used atomic targets Ar, Kr, and Xe as examples. Each atom has 8 electrons in the valence shell. The calculated energies of the highest occupied $n$p orbitals and the energies of the upper $(n+1)$s orbitals are summarized in the Supplementary material, where $n$ is the corresponding principal quantum number. We apply a trapezoidal driving laser pulse with a total duration of 8 optical cycles (with 2-cycle linear ramps and 4-cycle constant center). The laser wavelength is 800 nm and the laser intensity is $1\times10^{14}$ W/cm$^2$. The obtained harmonic spectra are shown in Fig. 1.  To have a better view of the results, the spectra are smoothed by convolution with a Gaussian function \cite{Lein2}. The smoothed harmonic spectra below the 15th harmonic order are presented in the insets. One can clearly see from the spectra that the HHG is substantially suppressed in particular harmonic orders in the below-threshold region. The suppressions locate around the 7th harmonic order for Ar and Kr, and the 5th harmonic order for Xe. The intensities of the suppressed harmonics are 2-3 orders of magnitude lower than those in the plateau region.

%
%
%
%

\begin{figure}[htb]
\centerline{
\includegraphics[width=5cm]{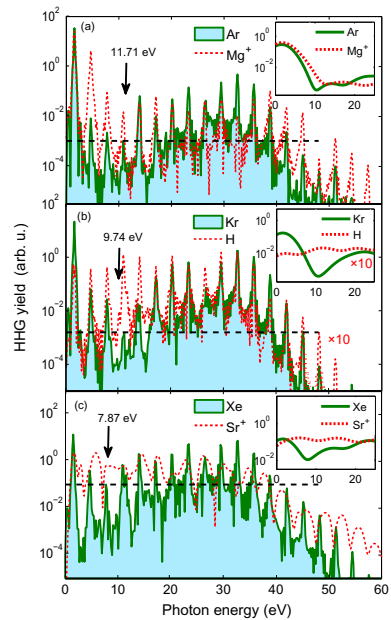}}
\caption{\label{Compare} High harmonic spectra from (a) Ar, (b) Kr, and (c) Xe and respect reference systems. The wavelength of the driving laser pulse is 800 nm and the laser intensity is $1\times10^{14}$ W/cm$^2$. The smoothed harmonic spectra below the 15th harmonic order are presented in the insets. The black arrows indicate the energy gaps between the $n$p and $(n+1)$s states, and the horizontal dashed lines are plotted to guide the eyes.
}
\end{figure}

To clarify the origin of this suppression, we repeat our calculations for the target ions and atom with only one electron in the valance shell and with similar ionization potentials for comparison. The reference systems are Mg$^+$ ($E_h=-15.15$ eV) for Ar, H ($E_h=-13.58$ eV) for Kr, and Sr$^+$ ($E_h=-10.93$ eV) for Xe, respectively. $E_h$ is the calculated energy of the highest occupied orbital for the reference system. The obtained harmonic spectra are presented in Fig. 1 as the dashed curves. In contrast to those for Ar, Kr, and Xe, the harmonic intensities in the below-threshold region are as high as those in the plateau region. No suppression occurs in the harmonic spectra from the reference systems. The comparison shows that the suppression effect is a result of multi-election dynamics. We have also calculated the high harmonic spectrum from other atoms in the third period of the periodic table from Mg to Cl with TDDFT. A general tendency is found that, the similar suppression effect becomes more obvious with the increase of number of electrons in the valence shell (see the Supplementary material).

\begin{figure}[htb]
\centerline{
\includegraphics[width=8.3cm]{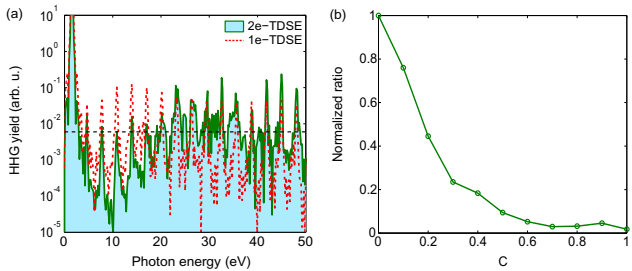}}
\caption{\label{TDSEfig} (a) High harmonic spectra obtained from one-dimensional two-electron TDSE (green curve) and one-electron TDSE (red curve). (b) Normalized ratio between the intensities of the 7th harmonic and the 29th-129th harmonics with $C$ varying from 0 to 1 in the two-electron TDSE simulation.}
\end{figure}

To further validate that the suppression effect results from the multi-electron dynamics, we also calculate the harmonic spectrum by numerically solving the exact one-dimensional two-electron TDSE
\begin{equation}\label{TDSEeq}
    i\frac{\partial}{\partial t}\Psi(x_1,x_2,t)=\hat{H}\Psi(x_1,x_2,t),
\end{equation}
with
\begin{equation}\label{TDSEeq}
    \hat{H}=\sum_{i=1}^2[-\frac{1}{2}\frac{\partial^2}{\partial x_i^2}+V_i(x_i)+E(t)x_i]+V_{ee}(x_1,x_2).
\end{equation}
$x_1$ and $x_2$ are the coordinates for the two electrons in the system. For a modeled He, soft-core potentials are applied $V_{i}=-2/\sqrt{\alpha^2+x_i^2}$, $V_{ee}=1/\sqrt{\alpha'^2+(x_1-x_2)^2}$. The soft core parameter $\alpha=\alpha'=0.5$ to fit the ground state energy of He. In this calculation, the laser intensity is $1\times10^{15}$ W/cm$^2$ and the other laser parameters are the same as those for Fig. \ref{Compare}. The obtained high harmonic spectrum is shown in Fig. \ref{TDSEfig}(a) as the green curve. It is found that, the harmonics around the 7th harmonic order are substantially suppressed. The intensities of the suppressed harmonics are 1-2 orders of magnitude lower than that in the plateau region. For comparison, the harmonic spectrum obtained by solving the one-electron TDSE with SAE soft-core potential is also shown as the dashed red curve in Fig. \ref{TDSEfig}(a). It is found that, there is no suppression observed in the SAE harmonic spectrum.

Moreover, in the two-electron TDSE simulation, we modify the $e$-$e$ interaction as $V_{ee}=C/\sqrt{\alpha'^2+(x_1-x_2)^2}$ to demonstrate the role of $e$-$e$ interaction on the suppression. The $e$-$e$ interaction is taken into account when $C=1$ and is ignored when $C=0$. We vary the coefficient $C$ from 0 to 1 and calculate the ratios between the intensities of the 7th harmonic and the 29th-129th harmonics in the plateau region. When $C$ is changed, the value of the soft-core parameter is also adjusted to keep the system energy remaining the same. The ratios normalized with the value at $C=0$ are shown in Fig. \ref{TDSEfig}(b). It is found that, with the increase of $e$-$e$ interaction (C varying from 0 to 1), the ratio decreases correspondingly. All the results obtained with TDSE simulation indicate that the suppression effect in the below-threshold HHG arises from the $e$-$e$ interaction.

\begin{figure}[htb]
\centerline{
\includegraphics[width=8.5cm]{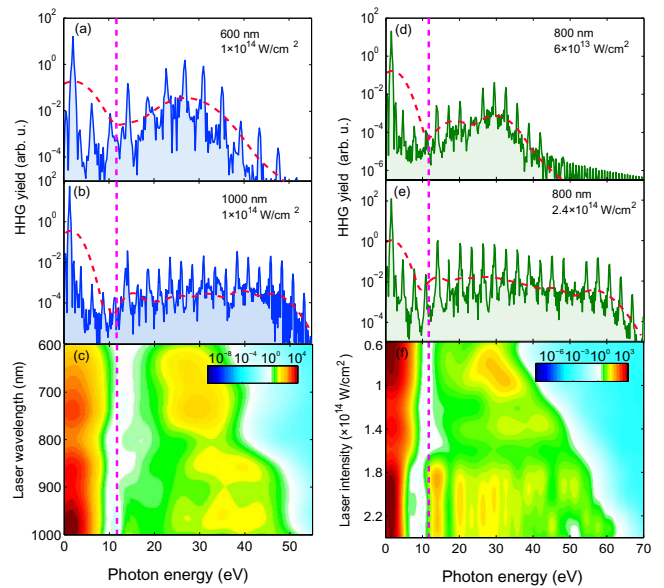}}
\caption{\label{Scan} High harmonic spectra from Ar by varying (a-c) the laser wavelength and (d-f) the laser intensity. Panels (a-b) and (d-e) show the harmonic spectra with particular laser parameters. The red dashed curves are the smoothed spectra. Panel (c) shows the smoothed harmonic spectra with the laser wavelength varying from 600 nm to 1000 nm. Panel (f) shows the smoothed harmonic spectra with the laser intensity varying from $6\times10^{13}$ W/cm$^2$ to $2.4\times10^{14}$ W/cm$^2$. The purple vertical lines indicate the energy gap between the 3p orbital and the 4s orbital of Ar.}
\end{figure}

The results shown in Fig. 1 reveals another important phenomenon: the photon energies of the most suppressed harmonics approximately equal the energy gaps between the highest occupied $n$p orbitals and the upper unoccupied $(n+1)$s orbitals. This is further examined by scanning the wavelength and intensity of the driving laser pulses. The results for Ar are shown in Fig. \ref{Scan}. The vertical lines indicate the value of the energy gap between the 3p orbital and the 4s orbital of Ar. It is shown that, the positions of the suppressions remain the same irrespective of the change of the laser parameters. We have also calculated the harmonic spectra for Kr and Xe with various laser wavelengths and intensities, and find that the positions of the suppressions remain unchanged (see the Supplementary material).

\begin{figure}[htb]
\centerline{
\includegraphics[width=8.6cm]{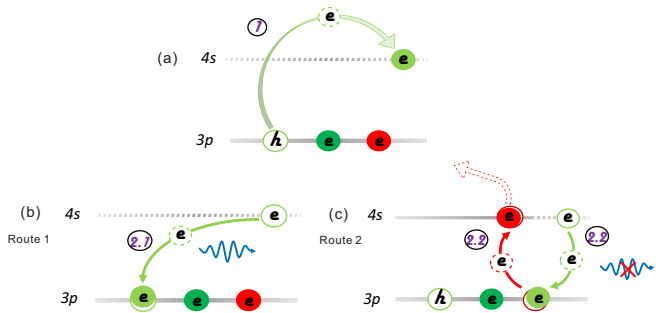}}
\caption{\label{Schematic} Schematic illustration of the quenching effect in HHG taking the target Ar as an example. (a) The considered electron is ionized and accelerated by the laser field, and finally recombines with the energy equal to that of the 4s orbital. This electron may recombine to the 3p orbital and release the accumulated energy in two routes. (b) In Route 1, it releases the energy in the form of a harmonic photon with energy $\Delta E$. (c) In Route 2, it transfers the energy to another electron. The second electron is excited to the 4s orbital and further ionized by the laser field. Therefore, no harmonic is generated in Route 2.}
\end{figure}

To reveal how the multi-electron dynamics result in the suppressions in the below-threshold region, we revisit the three-step model of HHG \cite{Corkum}. The schematic is shown in Fig. \ref{Schematic} taking the target Ar as an example. According to the three-step model, an HHG process is initialized by the ionization of electrons in the highest occupied orbital, i.e. the 3p orbital (Fig. \ref{Schematic}(a)). The electrons are then accelerated by the laser field and finally return to the core. Now we consider the return electron with the energy equal to that of the 4s orbital. In the recombination step, this electron may recombine to the 3p orbital and release the accumulated energy in two routes. In Route 1, it releases the energy by emitting a photon. This is the typical HHG process, where high harmonic emission with photon energy equal to the energy gap between 3p and 4s orbitals $\Delta E$ is obtained (Fig. \ref{Schematic}(b)). In Route 2, the considered electron returns to the ground state without emitting a harmonic photon. Alternatively, it transfers the energy to another electron (Fig. 4(c)). The second electron is excited to the 4s orbital and further ionized by the laser field \cite{NDSI}. The total amount of return electrons is constant. Since a portion of the electrons recombine via Route 2, the number of electrons recombining via Route 1 is decreased. As a result, the harmonics emitted via Route 1 (photon energy $\sim\Delta E$) are suppressed. This suppression effect is analog to the fluorescence quenching owing to the energy transfer between molecules in fluorescent materials. We therefore call it the quenching effect in HHG. To our knowledge, such a quenching effect in HHG has never been addressed before.

The suppression for the quenched harmonics depends on the probabilities of the radiative recombination (Route 1) and the non-radiative recombination (Route 2). The higher the probability of Route 2 is, the more the harmonics will be suppressed. As Route 2 is a resonant process, the probability of Route 2 can be much higher than that of Route 1, and thus the quenched harmonics can be suppressed by a few orders of magnitudes. Besides, the suppression also depends on the ionization rate of the higher-lying orbital. The ionization rate of the higher orbital should be larger than the rate of radiative transition back to the ground orbital. Otherwise, the second electron may return to the ground state via radiative transition, and then the quenching effect will be eliminated. One may even expect a resonant enhancement of HHG \cite{Xiong} if the ionization rate is much smaller than the radiative transition rate.



The crucial process for the quenching effect is the energy transfer when the first electron returns to the core. To verify this process, we consider the observable quantity of ellipticity dependence of HHG. The HHG efficiency decreases with the increase of laser ellipticity, because the ionized electron will drift in the transverses direction and fail to return to the core \cite{Corkum}. Meanwhile, the probability of the $e$-$e$ energy transfer process in Route 2 decreases and thus the quenching effect will vanish with the increase of the ellipticity. Based on this effect, it is expected that the intensities of the quenched harmonics will decrease more slowly (even increase) with the increase of ellipticity.

\begin{figure}[htb]
\centerline{
\includegraphics[width=8cm]{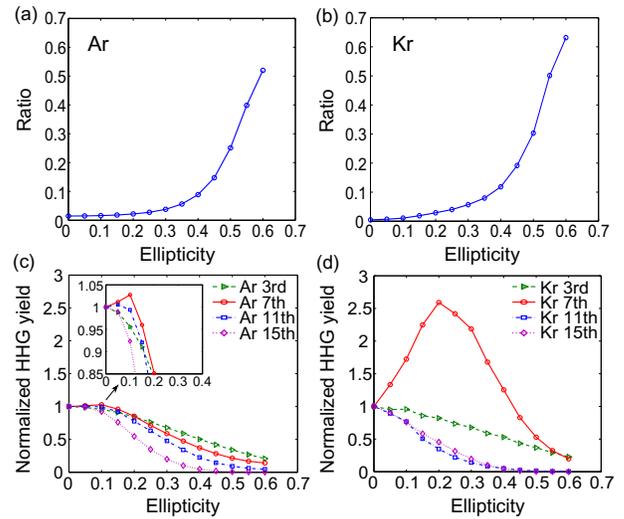}}
\caption{\label{ellipticity} (a-b) Ratios between the intensities of the 7th harmonic and the 11th-15th harmonics with various ellipticities for Ar and Kr, respectively. (c-d) Normalized intensities of the 3rd, 7th, 11th and 15th harmonics with the increase of ellipticity for Ar and Kr, respectively.}
\end{figure}

To demonstrate this effect, we calculate the harmonic spectra from Ar and Kr driven by laser fields with different ellipticities. The laser intensity and wavelength are the same as those for Fig. 1. Figs. \ref{ellipticity}(a) and (b) show the ratios between the intensities of the 7th harmonic and the 11th-15th harmonics for Ar and Kr, respectively. For both targets, the ratios increase with the ellipticity. This indicates that the suppression at the 7th harmonic becomes weaker with the increase of the ellipticity. Figs. \ref{ellipticity}(c) and (d) show the normalized intensities of the 3rd, 7th, 11th and 15th harmonics for Ar and Kr, respectively. Anomalous ellipticity dependence at the quenched 7th harmonic is found in both Figs. \ref{ellipticity}(c) and (d): the intensities of the 7th harmonics increase first when the ellipticity is moderate and then drop with the increase of ellipticity, while the intensities of other harmonics decrease with the increase of ellipticity as normal. Comparing Figs. \ref{ellipticity}(c) and (d), it is also found that the anomalous ellipticity dependence is more obvious for Kr than for Ar. This is consistent with the fact in Fig. \ref{Compare} that the quenching effect is much stronger for Kr than for Ar. All the results for the ellipticity dependence verify that the quenching effect is due to the multi-electron dynamics involving the $e$-$e$ energy transfer.

In conclusion, we theoretically demonstrate the quenching effect in below-threshold HHG with the TDDFT and TDSE simulation. The commonly used multi-electron targets Ar, Kr, Xe are discussed. It is found that, the HHG at particular harmonic orders is substantially suppressed when multi-electron interaction comes into play. The intensities of the quenched harmonics are 2-3 orders of magnitude lower than those in the plateau regions. It is also shown that the positions where the quenching occurs are determined by the energy gaps between the highest occupied orbitals and the higher orbitals. The results are attributed to a new class of dynamics involving the $e$-$e$ energy transfer. The dynamics is further verified from the anomalous ellipticity dependence of the quenched harmonics. This work is the first report of the quenching effect in HHG, which reveals the important role of the dynamical multi-electron interaction on the HHG process especially in the below-threshold region. The study will help people gain a deeper insight into the HHG process.

\section*{Acknowledgment}
We thank Prof. Zengxiu Zhao, Prof. Manfred Lein, and Prof. Jing Chen for invaluable discussions. This work was supported by the National Natural Science Foundation of China under Grants No. 11234004, No. 11404123 and No. 61275126, and the 973 Program of China under Grant No. 2011CB808103. Numerical simulations presented in this paper were carried out using the High Performance Computing Center experimental testbed in SCTS/CGCL (see http://grid.hust.edu.cn/hpcc).

\end{document}